# Realization of valley-controlled orbital angular momentum via inter-orbital coupling

Wenxia Xu[1†], Ziteng Wang[1†], Yihan Wang[1†], Qinggang Shang[1], Liqin Tang[1], Jingjun Xu[1], Hrvoje Buljan[1,3], Daohong Song[1,2*] and Zhigang Chen[1,2*]

[1]*The MOE Key Laboratory of Weak-Light Nonlinear Photonics, TEDA Applied Physics Institute and School of Physics, Nankai University, Tianjin 300457, China*
[2]*Collaborative Innovation Center of Extreme Optics, Shanxi University, Taiyuan, Shanxi 030006, China*
[3]*Department of Physics, Faculty of Science, University of Zagreb, Bijenička c. 32, Zagreb 10000, Croatia*
[†]*These authors contributed equally to this work*
[*]*songdaohong@nankai.edu.cn, zgchen@nankai.edu.cn*

**Abstract:**

Orbital angular momentum (OAM) represents a fundamental degree of freedom (DoF) for understanding novel phenomena in solids and synthetic lattices. However, the generation and control of OAM in lattice systems remain challenging. Here, we experimentally observe valley-controlled OAM states in inversion-symmetry-broken photonic honeycomb lattices consisting of *s*- and higher-orbital sublattices. These OAM states manifest as chiral vortex arrays in real space. By selectively exciting *s*-orbital modes carrying no initial OAM at either the $K$ or $K'$ valley, we observe higher-orbital OAM modes generated via inter-orbital coupling in neighboring sites. Specifically, *s*-*p* orbital coupling generates OAM states with topological charge $l = \pm 1$, while *s*-*d* orbital coupling yields higher-charge OAM states with $l = \pm 2$. The sign of the topological charge is determined by the valley DoF, enabling deterministic control of vortex chirality across the entire array. Our theory further reveals that a local vortex-chirality symmetry enforces the valley-dependent selection of vortex chirality at the $K$ and $K'$ valleys, establishing a direct connection between the valley DoF and real-space vortex chirality. These results establish a versatile platform for engineering chiral vortex arrays via orbital physics and exploring orbital- and topology-mediated phenomena in photonic systems and beyond.

*Introduction* — Orbital degree of freedom (DoF) plays a significant role in condensed matter physics, underpinning a wide range of phenomena from unconventional magnetism and high-temperature superconductivity to topological states [1–4]. In addition to charge and spin, orbital DoF describes the spatial distribution of electron wavefunctions. In atoms, continuous rotational symmetry allows complex orbitals to carry well-defined orbital angular momentum (OAM). However, in solids, it is commonly believed that crystal potentials typically quench OAM by breaking continuous rotational symmetry and imposing specific symmetries on Bloch states [5]. This conventional wisdom has been challenged by recent developments in the emerging field of orbitronics, where OAM can be generated and transported in materials with broken inversion symmetry or under external electric fields, even in the absence of spin–orbit coupling [6–9]. These developments have established OAM as an active DoF in solids, playing a fundamental role in exotic phenomena such as the orbital-Hall effect [6,8,10,11], the spin-Hall and valley-Hall effects [6,12–14]. Despite these advances, the generation and direct observation of OAM in solids remain challenging, as orbital dynamics are typically entangled with spin and other DoF [9,11,12,15,16].

Synthetic orbital lattices provide a versatile platform to explore orbital-mediated phenomena in solids and have attracted considerable attention in various physical systems, including photonic lattices and photonic crystals [17–23], optical lattices for ultracold atoms [24–26], and engineered electronic lattices [27,28]. Importantly, the parameters governing the orbital-band structure in synthetic lattices can be flexibly engineered, even in regimes inaccessible to natural solids, thereby enabling the exploration of orbital physics difficult to access in condensed matter systems. Indeed, a variety of unconventional orbital-mediated phenomena have been experimentally realized in synthetic lattices, including orbital superfluidity [24–26], topological boundary states [19,29], tilted Dirac cones [30], and orbital-induced synthetic flux [29,31]. However, despite advances in artificial orbital lattices, most work has focused on orbital effects without OAM. Recently, it has been shown that OAM ordering with spontaneous time-reversal symmetry breaking can emerge in *p*-orbital optical lattices of cold atoms, where its signature was identified in momentum space [25,26]. Thus far, controllable generation and direct real-space observation of OAM in synthetic orbital lattices remain elusive.

In this work, we experimentally demonstrate valley-controlled generation of OAM states in an inversion-symmetry-broken photonic honeycomb lattice (HCL) via inter-orbital coupling. The HCL is optically induced in a nonlinear crystal with tailored on-site potentials for the two sublattices, such that one sublattice hosts *s* orbitals whereas the other supports higher-order orbitals. Under valley-selective excitation, engineered inter-orbital coupling converts *s*-orbital excitations into OAM states in the higher-orbital sublattice, yielding real-space vortex arrays with valley-dependent chirality. When the *s*-*p* orbital coupling dominates, excitation of $s$-orbital states at the $K$ ($K'$) valley generates OAM states with topological charge (TC) $l = 1$ ($l = -1$) in $p$-orbital sublattice [Fig. 1(a)]. For the HCL with dominant *s*-*d* coupling, OAM states emerge as doubly-charged optical vortex arrays. Theoretically, we show that the vortex chirality is dictated by a local *vortex-chirality symmetry* (VCS) at the high-symmetry $K$ ($K'$) points, while the inter-orbital conversion undergoes propagation-dependent oscillations owing to inter-band mode beating. Compared with natural HCL-structured monolayer transition metal dichalcogenides (TMDs), our platform allows dynamic tuning of the *s*-*p* and *s*-*d* orbital hybridization via optical writing, enabling the exploration of real-space propagation dynamics including inter-orbital Rabi-like power beating, and macroscopic phase singularities localized at individual lattice sites. These capabilities go beyond the momentum-space valley pseudospins characteristic of natural TMDs.

We consider a photonic HCL with inequivalent sites (A and B) per unit cell [Fig. 1(a)]. The A sublattice sites have a higher refractive index and support $p$-orbital modes, while the B sublattice sites feature a lower refractive index and host only $s$-orbital modes. The inter-orbital coupling between the $s$-orbitals in B sublattice and $p$-orbitals in A sublattice can be achieved by tuning the effective refractive-index contrast between the two sublattices [17]. The twofold degenerate $p$-orbital modes ($p_x$ and $p_y$) at the A sites can form OAM modes with opposite chirality through superpositions $+l = p_x + ip_y$ and $-l = p_x - ip_y$, corresponding to vortex modes with TCs $l = 1$ and $l = -1$, respectively. For the *s*-*p* orbital coupling case, an input $s$-orbital excitation at the $K$ point of the first Brillouin zone (BZ) generates OAM states in the A sublattice, manifesting as chiral vortex arrays with TC $l = +1$ [Fig. 1(b), left], whereas excitation at the $K'$ point yields vortex arrays with synchronized TC $l = -1$ [Fig. 1(b), right]. The selection rules for valley-dependent vortex chirality ($+1$, –1) are summarized in Fig. 1(b).

The underlying physics of the HCL with *s*-*p* orbital coupling can be described by a tight-binding model in the $[s, +l, -l]^T$ orbital basis [26]. The momentum-space Bloch Hamiltonian is:

$$H(\boldsymbol{k}) = \begin{bmatrix} \varepsilon_s & h_{12}(\boldsymbol{k}) & h_{13}(\boldsymbol{k}) \\ h_{12}^*(\boldsymbol{k}) & \varepsilon_v & 0 \\ h_{13}^*(\boldsymbol{k}) & 0 & \varepsilon_v \end{bmatrix}, \tag{1}$$

with $h_{12} = t_1 e^{i\boldsymbol{k}\cdot\boldsymbol{e}_1} + t_2 e^{i\boldsymbol{k}\cdot\boldsymbol{e}_2} + t_3 e^{i\boldsymbol{k}\cdot\boldsymbol{e}_3}$ and $h_{13} = t_1^* e^{i\boldsymbol{k}\cdot\boldsymbol{e}_1} + t_2^* e^{i\boldsymbol{k}\cdot\boldsymbol{e}_2} + t_3^* e^{i\boldsymbol{k}\cdot\boldsymbol{e}_3}$, where $\varepsilon_s$ denotes the on-site propagation constant of the $s$-orbital mode in the B sublattice, while $\varepsilon_v$ denotes the degenerate on-site propagation constant of the two vortex modes $|+l\rangle$ and $|-l\rangle$ in the A sublattice. The nearest-neighbor vectors are $\boldsymbol{e}_1 = a(\sqrt{3}/2, 1/2)$, $\boldsymbol{e}_2 = a(-\sqrt{3}/2, 1/2)$, $\boldsymbol{e}_3 = a(0, -1)$, with $a$ being the lattice constant.

The inter-orbital coupling coefficients in $H(\boldsymbol{k})$ are $t_1 = t$, $t_2 = te^{i|l|2\pi/3}$ and $t_3 = te^{-i|l|2\pi/3}$. For the $l = 1$ vortex modes, this reduces to $t_1 = t$, $t_2 = te^{i2\pi/3}$ and $t_3 = te^{-i2\pi/3}$. Details for Hamiltonian transformation from *s*-*p* to *s*-vortex orbital basis are provided in Supplementary Information (SI). Diagonalizing $H(\boldsymbol{k})$ gives the band structure $\beta(\boldsymbol{k})$. For the resonant case with $t = 1, a = 1,$ and $\varepsilon_s = \varepsilon_v = 1$, the band structure [Fig. 1(c)] contains two dispersive bands separated by a central flat band. The three bands touch at the $\Gamma$ point, while they are gapped at $K$ and $K'$ points in the first BZ. The central flat band originates from destructive interference between the two vortex-orbital coupling channels, which gives an orbital combination decoupled from the *s* mode. At the $\Gamma$ point, $t_1 + t_2 + t_3 = 0$, so that $h_{12} = h_{13} = 0$. Together with the resonant condition $\varepsilon_s = \varepsilon_v$, this gives the threefold degeneracy shown in Fig. 1(c). Figure 1(c) also shows eigenmodes of the bottom dispersive band within a hexagonal unit cell at $K$ and $K'$ points. Each *p*-orbital site in A sublattice supports a vortex mode with TC $l = +1$ $(-1)$ at $K$ $(K')$ point across the entire lattice. Notably, $\varepsilon_s = \varepsilon_v$ is the ideal resonance condition, under which the *s* modes and vortex modes can undergo complete conversion. When $\varepsilon_s$ and $\varepsilon_v$ are slightly detuned, vortex generation in the A sublattice persists, but the conversion efficiency decreases as the detuning increases. The degenerate vortex states at $K$ and $K'$ points offer the valley DoF that is related by time-reversal symmetry [32]. Thus, no net OAM exists if the HCL preserves time-reversal symmetry, since the vortex modes with opposite TCs at $K$ and $K'$ valleys coexist and cancel each other. Interestingly, OAM states with TC $|l| = 1$ have been experimentally demonstrated in momentum space via nonlinearity-induced spontaneous time-reversal symmetry breaking in cold-atom systems. However, these OAM states emerge spontaneously, making their chirality difficult to control [26]. In what follows, we demonstrate controllable real-space OAM states in photonic HCLs without breaking the time-reversal symmetry, but rather by selectively exciting the $K/K'$ valleys in momentum space through inter-orbital coupling.

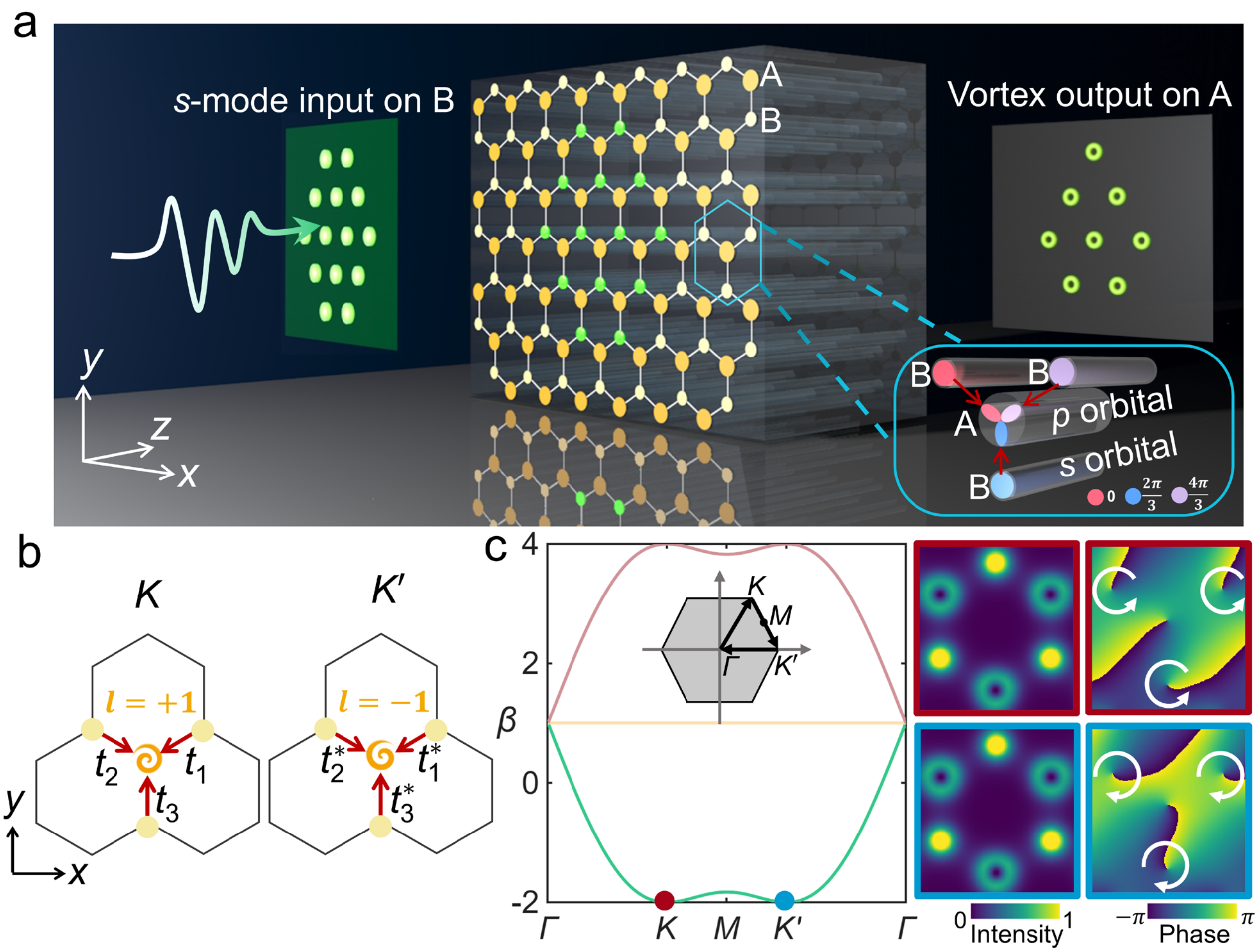

**Fig. 1: Inter-orbital-coupling-enabled generation of OAM states.** (a) Schematic of OAM state generation in a symmetry-broken photonic HCL composed of *p*-orbital A sublattice and *s*-orbital B sublattice. The bottom-right inset illustrates that the inter-orbital coupling converts the *s*-orbital excitation from the B sublattice into *p*-orbital OAM states in the A sublattice. (b) Valley-dependent vortex chirality ($l = +1$ or $l = -1$) determined by the high-symmetry points ($K$ or $K'$) in momentum space. For the *s*-*p* coupling case, the A sublattice hosts vortices with TC $l = \pm 1$, where the sign depends on the high-symmetry point ($l = +1$ at $K$ point, and $l = -1$ at $K'$ point). (c) Left: tight-binding band structure along the high-symmetry points, showing two dispersive bands (red and cyan) and a flat band (yellow). Red and blue dots mark $K$ and $K'$ points, respectively. Inset: the first BZ. Right: intensity (left column) and phase (right column) distributions of Bloch modes at $K$ (top row) and $K'$ (bottom row) points. White circular arrows indicate the vortex chirality.

First, we experimentally demonstrate the generation of OAM states in a photonic HCL via *s*-*p* orbital coupling. The *s*-*p* orbital HCL is optically induced in a nonlinear crystal [33,34] (see SI for details). The induced HCL [Fig. 2(a)] has two sublattices with unequal refractive indices, where the A sublattice has a higher refractive index to support $p$-orbital modes, while the B sublattice with a lower refractive index only hosts $s$-orbital modes. Efficient $s$-$p$ orbital coupling is controlled by tuning the lattice spacing and the relative refractive-index contrast of the two sublattices. In our experiment, the HCL-writing beam with a wavelength of 532 nm and an input power of 0.8 mW was launched into the crystal under a bias voltage of 900 V. The resulting lattice spacing is about 35 μm, and the refractive-index contrasts of the induced $s$-type and $p$-type waveguides are approximately $\Delta n_s = 1.98 \times 10^{-4}$ and $\Delta n_p = 2.97 \times 10^{-4}$, respectively. The probe beam is a truncated triangle-lattice beam, constructed by interfering three broad Gaussian beams matched to three $K$ points in momentum space, in analogy to the pseudospin excitation in photonic graphene [33,35–37]. The input probe beam covers the B sublattice to excite the $s$-orbital state [Fig. 2(f)]. After 15mm of propagation through the lattice, the probe beam couples to the $p$-orbital modes with doughnut-like ring patterns on each A-sublattice site [Fig. 2(b) and Fig. 2(c)]. To reveal the phase structures of the $p$-orbital modes, an inclined broad beam is introduced to interfere with the output. The fringe bifurcation shows that the ring-like patterns in the A sublattice form a vortex array with the same TC $l = +1$ [Fig. 2(d)]. Although the probe beam initially excites the three $K$ valleys in the first BZ [Fig. 2(e)], the output spectrum exhibits more pronounced higher-order Bragg peaks consistent with the excitation of $p$-orbital vortex modes. These results confirm that OAM modes with TC $l = +1$ are generated via $s$-$p$ orbital coupling, in agreement with the theoretical Bloch mode at $K$ points shown in Fig. 1(c).

Interestingly, the chirality of the OAM states depends on selective excitation of $K$ or $K'$ valley in momentum space. When the probe beam [Fig. 2(f)], whose spectrum matches the three $K'$ points [the red circles in Fig. 2(j)], excites $s$-orbital states in the B sublattice at input, the output intensity and interferogram [Fig. 2(h) and 2(i)] indicate that a vortex array with TC $l = -1$ emerges in the A sublattice [Fig. 2(g)]. The experiments agree well with the theoretical Bloch modes at $K'$ points [Fig. 1(c)] and simulations in SI. For comparison, if the probe beam excites $s$-orbital modes on A sublattice, the output remains localized at the A sublattice, showing an $s$-orbital mode profile and negligible coupling to the B sublattice. This is because the propagation constant of $s$-orbital mode in A-sublattice waveguide does not match that of any modes in B-sublattice waveguide. The corresponding experimental results are shown in SI. This again confirms that OAM states originate from the inter-orbital coupling between $s$ orbital in B sublattice and $p$ orbital in A sublattice.

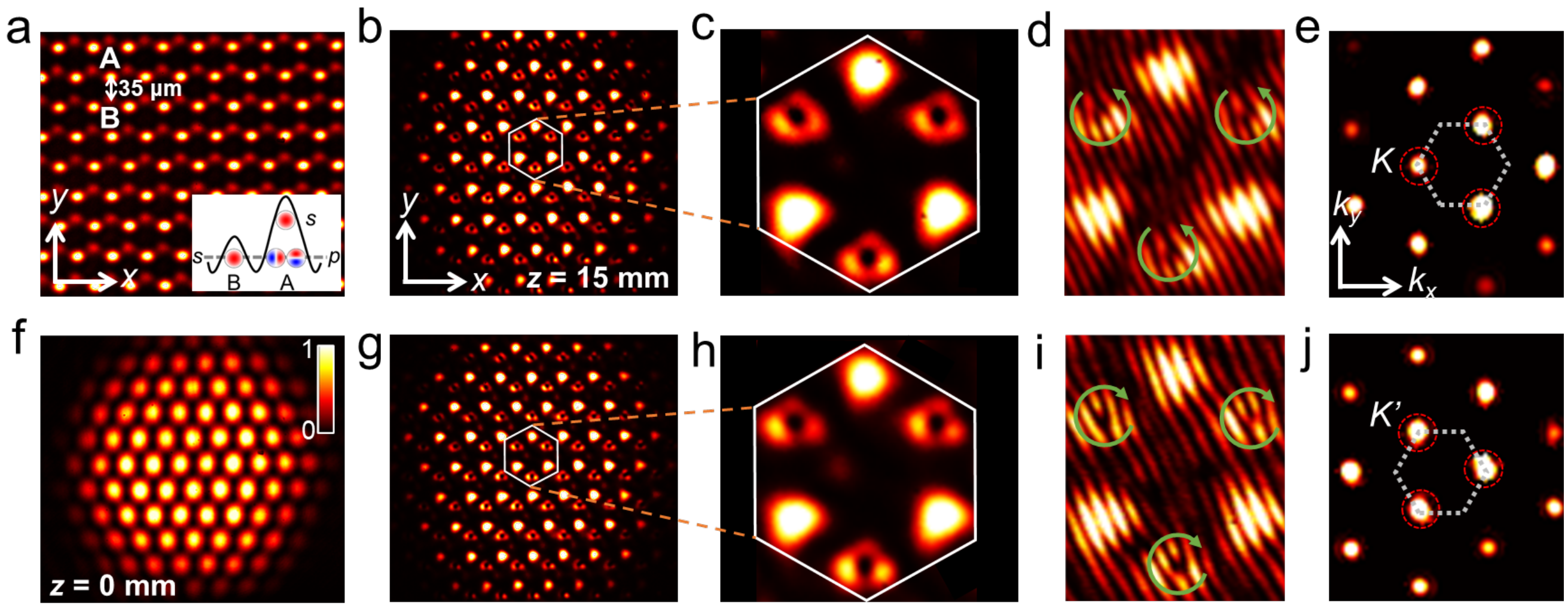

**Fig. 2: Experimental demonstration of valley-dependent OAM states through *s*-*p* orbital coupling.** (a) Experimentally induced photonic *s*-*p* orbital HCL. The bottom-right inset shows the *s*-*p* orbital coupling where the *s* orbital in B sites and the *p* orbital in A sites have matched propagation constants. (f) Input intensity pattern of the probe beam at $z = 0$ for excitation of *s* orbital in B sublattice. (b-e) Experimental results for excitation of the three $K$ valleys in momentum space. (b) Output intensity pattern after a propagation distance of $z = 15$mm. (c) Magnified view of the boxed region in (b). (d) Output interferogram corresponding to (c), used to identify the vortex phase singularities. The green curved arrows indicate the position and chirality of vortices at A sublattice sites. (e) Fourier spectrum of the output, where the white dashed hexagon marks the first BZ and the red dashed circles highlight the initially excited three $K$ valleys. (g-j) Experimental results corresponding to (b-e), but for the excitation of three $K'$ valleys in momentum space.

To elucidate the underlying physics of valley-dependent OAM states, we introduce the concept of local VCS: it resembles chiral (sublattice) symmetry [38], is defined for the vortex-mode basis, and links synchronized vorticity to the high-symmetry points in momentum space (the valley-selection rule). In the $[s, +l, -l]^T$ orbital basis, the VCS operators $R_\pm$ are defined as

$$R_+ = \begin{bmatrix} -1 & 0 & 0 \\ 0 & 1 & 0 \\ 0 & 0 & -1 \end{bmatrix}, \qquad R_- = \begin{bmatrix} -1 & 0 & 0 \\ 0 & -1 & 0 \\ 0 & 0 & 1 \end{bmatrix}. \tag{2}$$

The symmetry holds when the Bloch Hamiltonian $H(k)$ in Eq. (1) commutes with the VCS operators $[H, R_\pm] = 0$, that is $R_\pm^{-1} H(k) R_\pm - H(k) = 0$. This condition holds only at special high-symmetry points in the BZ ($K$ or $K'$point). We quantify this condition using the Frobenius norm deviation as

$$\Omega_\pm(k) = \sqrt{\sum_{i,j} |[R_\pm^{-1} H(k) R_\pm - H(k)]_{ij}|^2}. \tag{3}$$

When $\Omega_+$ (or $\Omega_-$) vanishes at a given valley, the corresponding local VCS holds, as the Hamiltonian $H(k)$ and the symmetry operators $R_+$ (or $R_-$) share the same eigenstates. Note that $R_+$ (or $R_-$) isolates the $+l$ (or $-l$) vortex component in its eigenstates, i.e., $[0,1,0]^T$ is an eigenstate of $R_+$ (in the vortex-mode basis $[s,+l,-l]^T$) so the $s$-orbital cannot couple to the $+l$ (or $-l$) orbital at the corresponding valley. Figure 3 shows contour maps of $\Omega_\pm$ over the first BZ. For $|l| = 1$, $\Omega_+$ vanishes at three $K'$ points whereas $\Omega_-$ vanishes at three $K$ points [Figs. 3(a) and 3(b)], thereby identifying opposite vortex chirality at the two valleys ($K$ and $K'$points). Under excitation of the $s$-orbital in the B sublattice at $K$ points, OAM states with synchronized TC $l = +1$ are generated in the A sublattice. By contrast, the OAM states with $l = -1$ are suppressed given that $\Omega_-$ vanishes at $K$ points [Fig. 3(b)]. The local VCS requires the eigenstates of the Hamiltonian $H(k)$ at $K$ points to be eigenstates of $R_-$, so that the coupling of $s$-orbital to the $l = -1$ orbital state is forbidden. The same argument applies to OAM states with synchronized TC $l = -1$ [Fig. 3(a)]. The local VCS also applies to valley-dependent higher-order OAM generation. For example, we obtain OAM states with TCs $l = -2$ [Fig. 3(c)] and $l = +2$ [Fig. 3(d)] via $s$-$d$ orbital coupling, which also exhibits valley-dependent chirality at $K$ and $K'$ points.

Due to the limited propagation distance in our experiment, we numerically simulate the propagation dynamics of $s$-$p$ orbital coupling for much longer distances [Fig. 3(e)]. When the $s$-orbital B sublattice is initially excited, the power from the $s$ modes transfers to the vortex modes in A sublattice during propagation. The vortex-mode power peaks at $z = 2.61$cm. After that, the vortex-mode power then returns to the $s$ modes, with complete transfer at $z = 5.37$cm after one oscillation period. Real-space snapshots (insets) at $z = 8.08$cm and $10.78$cm show the periodic conversion between the $s$ modes and the vortex modes. This periodic oscillation arises from the beating between the modes of the top and bottom dispersive bands, with a period determined by their Bloch-mode propagation-constant difference at $K$ points (see SI for detailed theoretical analyses). For a selected initial valley, the vortex-array TCs remain invariant during propagation.

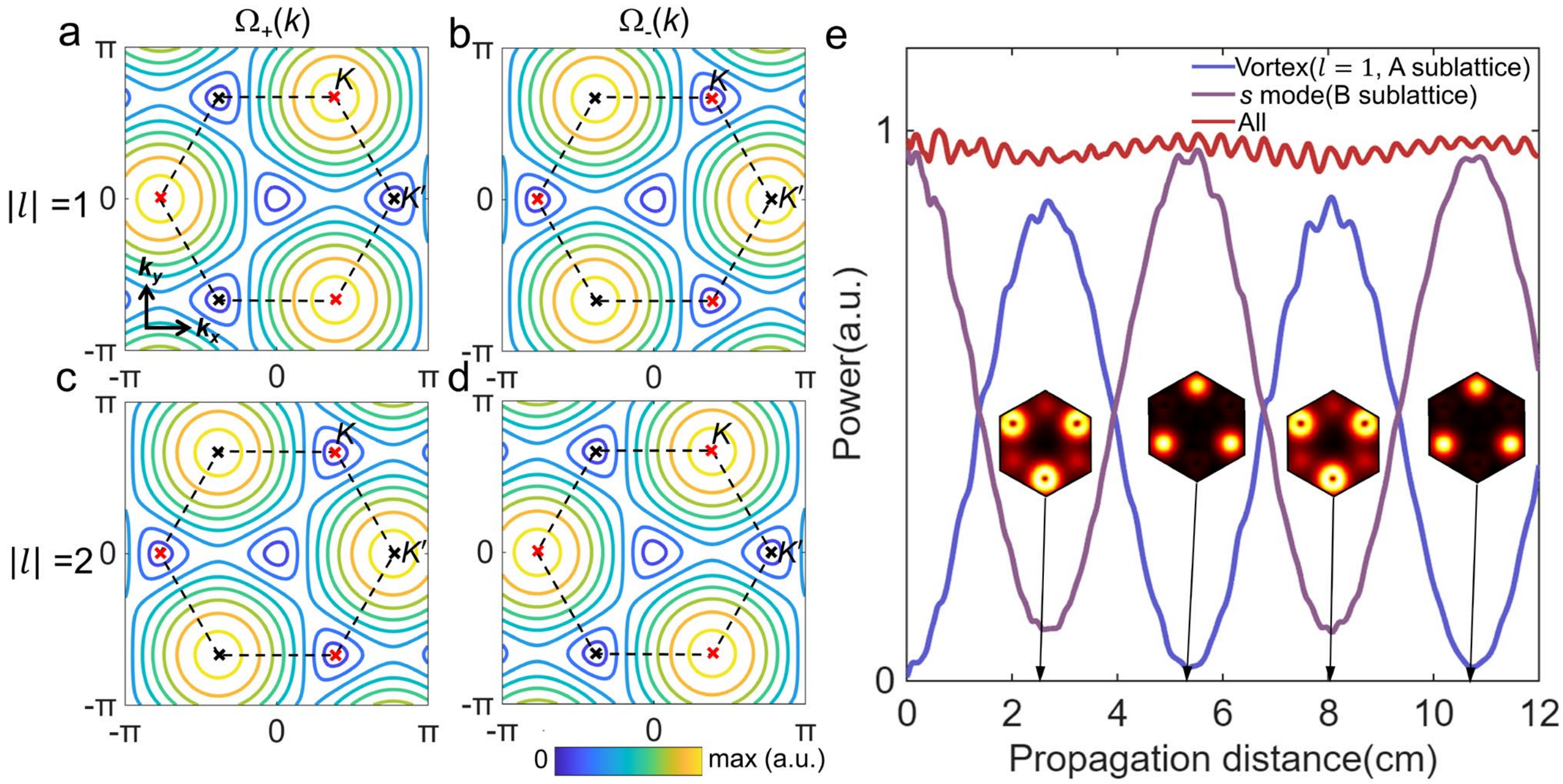

**Fig. 3: Theoretical analysis of valley-dependent OAM states based on the local VCS and long-distance propagation dynamics.** (a-d) Contour maps of $\Omega_{\pm}(k)$ over the BZ. Red (black) crosses denote the positions of $K$ ($K'$) points. The black dashed hexagon outlines the first BZ. (a, b) For the OAM states with $|l| = 1$, $\Omega_+$ vanishes at $K'$ points while $\Omega_-$ vanishes at $K$ points, thereby selecting opposite vortex chirality at the two valleys. (c, d) For the OAM states with $|l| = 2$, the selection rule is reversed between the $K$ and $K'$ points. (e) Evolution of modal power exchange between vortex mode (blue) and *s*-orbital mode (purple) during long-distance propagation obtained from numerical simulation, while the total power (red) is approximately conserved. Insets depict real-space snapshots at different propagation distances, illustrating periodic oscillation of power distribution between *s*-orbital mode on B sublattice and vortex mode on A sublattice.

Finally, we demonstrate the generation of OAM states with higher TC $l = -2$ via *s*-*d* orbital coupling. Figure 4(a) shows the optically induced photonic HCL with inequivalent A and B sublattices designed for dominant *s*-*d* coupling. For the induced *s*-*d* orbital HCL, the input power of the lattice writing beam is about 1.2 mW, and the bias voltage is 1150 V. The resulting lattice spacing is about 40μm, and the refractive indices for the *s*-orbital waveguides (B sublattice) and *d*-orbital waveguides (A sublattice) are $\Delta n_s = 1.36 \times 10^{-4}$ and $\Delta n_d = 2.97 \times 10^{-4}$, respectively. The input beam remains a truncated triangle-lattice beam [Fig. 4(f)] incident on the *s*-orbital B sublattice, with its spectrum occupying the three $K$ points in momentum space. After 15mm of propagation, the output intensity pattern exhibits a higher-order vortex array in the A sublattice [Fig. 4(b)] due to the *s*-*d* orbital coupling. The magnified interferogram in one hexagonal region reveals double-fork fringes, confirming a TC of $l = -2$ [Fig. 4(c) and 4(d)]. The output spectrum retains the initially excited three $K$ points in the first BZ [Fig. 4(e)], but Bragg peaks in higher-order BZs are also excited consistent with the excitation of the *d*-orbital modes in A sublattice. Numerical simulations corroborate these experimental results [Fig. 4(g-j)]. Notably, the experimental output [Fig. 4(b)] exhibits an imperfect intensity profile that deviates from the simulation [Fig. 4(g)], because the induced waveguides are not perfectly circular under strong anisotropic photorefractive nonlinearity [39]. Nevertheless, these experimental results agree with the theoretical Bloch modes at the $K$ point for *s*-*d* coupling, as shown in SI. Moreover, the OAM states with synchronized TC $l = +2$ can be generated by excitation of the three $K'$ points, as detailed in SI. These valley-dependent OAM states are also consistent with the local VCS shown in Fig. 3(c) and 3(d).

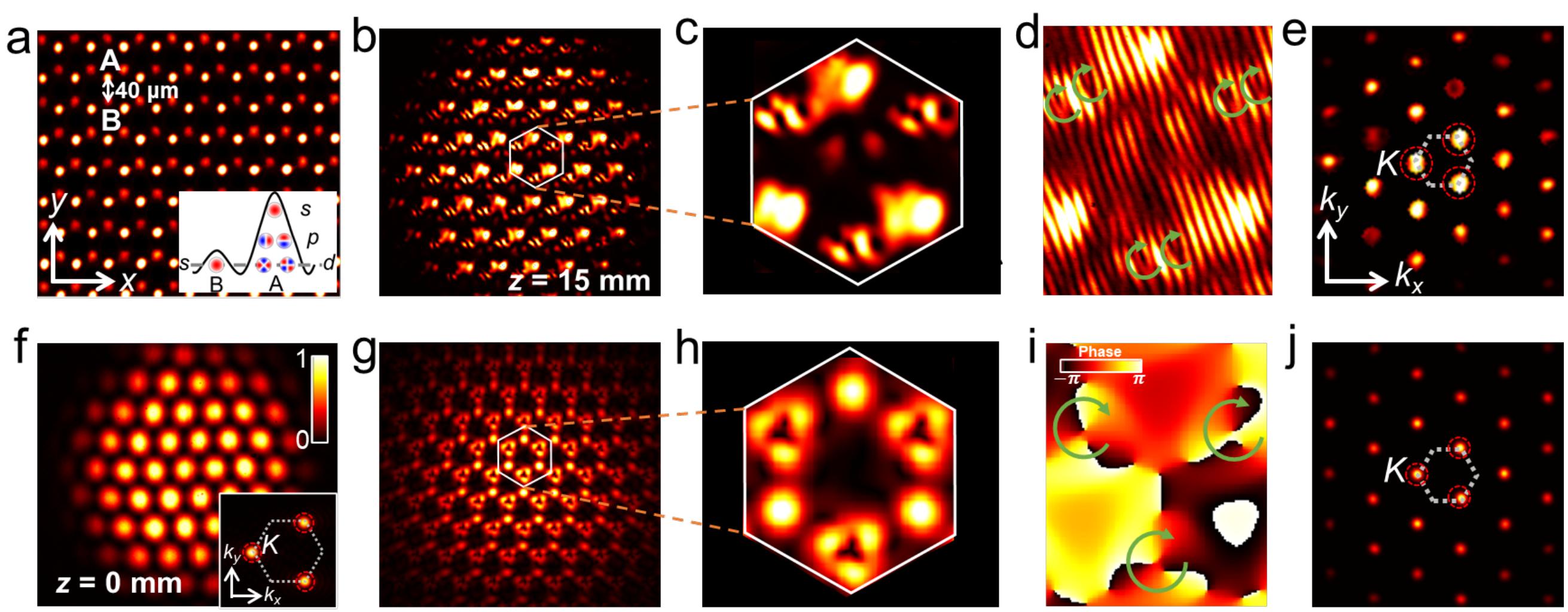

**Fig. 4: Experimental demonstration of OAM states with higher topological charge through *s*-*d* orbital coupling.** (a) An optically induced photonic *s*-*d* orbital HCL. The inset shows *s*-*d* orbital coupling between *s*-orbital in B sites and *d* orbital in A sites. (f) Input intensity pattern of the probe beam at $z = 0$. Inset: Fourier spectrum of the probe beam. (b-e) Experimental output at $z = 15\text{mm}$: (b) Output intensity pattern; (c) Magnified view of the hexagonal region highlighted in (b); (d) Output interferogram; the green curved arrows mark the positions and chirality of the vortices with topological charge $l = -2$; (e) Fourier spectrum of the output; the white dashed hexagon denotes the first BZ and red circles mark the initially excited three $K$ valleys. (g-j) Simulation results corresponding to (b-e).

In summary, we have experimentally demonstrated valley-controlled generation of OAM states through inter-orbital coupling in inversion-symmetry-broken photonic HCLs. Valley-dependent OAM states with $l = \pm 1$ and $l = \pm 2$ are realized through *s*-*p* and *s*-*d* orbital couplings, respectively, confirmed by real-space interference measurements and numerical simulations. We further show that the valley-dependent vortex chirality is determined by a local VCS in momentum space, establishing a direct connection between momentum-space symmetries and real-space vortex structures. Our results provide a general route to generate and control OAM states in synthetic orbital lattices and reveal a new mechanism for engineering chiral vortex arrays via orbital physics. The approach may also enable conversion between OAM states with different TCs. For example, by tuning relative on-site potentials so the relevant $p$- and $d$-orbital modes approach resonance, conversion between different OAM states ($|l| = 1$ and $|l| = 2$) may be realized.

Moreover, our results reveal a fundamental interplay among orbital DoF, valley physics, and real-space OAM structures in synthetic lattices, opening new avenues for exotic photonic band engineering and facilitating studies of intriguing topological phenomena [40-43]. Beyond the photonic HCL studied here, the underlying concept of inter-orbital OAM generation can be extended to a broad class of synthetic lattice platforms, including Kagome-type systems supporting nonlinear band-touching physics [35], super-honeycomb lattices with coexisting half-integer and integer pseudospins, and other synthetic lattices possessing engineered orbital DoF, nontrivial pseudospin structures, or hierarchical topological spectra. From an optical-engineering perspective, our photonic platform demonstrates valley-selective array-level OAM generation and conversion which may find applications in structured-light generation, mode-division multiplexing, and optical trapping and manipulation.

**Acknowledgments**

This research was supported by the National Natural Science Foundation of China (No. 124B2078, W2541003, 12625416，12274242, 12134006, and 12374309); the National Key R&D Program of China (No. 2022YFA1404800, 2025YFA1411800); the Fundamental Research Funds for the Central Universities (63263124) and the 111 Project (No. B23045) in China. H.B. acknowledges support from the project “Implementation of cutting-edge research and its application as part of the Scientific Center of Excellence for Quantum and Complex Systems and Representations of Lie Algebras,” (Grant No. PK.1.1.10.0004), co-financed by the European Union through the European Regional Development Fund - Competitiveness and Cohesion Programme 2021-2027.